\documentclass[aps,prb,preprint,showpacs,showkeys,floatfix]{revtex4}

\usepackage{graphicx,color}
\usepackage{multirow,slashbox}
\usepackage{amsmath}
\usepackage{cases}

\graphicspath{{figs/}}
\begin{document}

\title{A physical description for the monoclinic phase of zirconia based on the Tersoff potential}

\author{Jin-Wu Jiang}
    \altaffiliation{Corresponding author: jiangjinwu@shu.edu.cn, jwjiang5918@hotmail.com}
    \affiliation{Shanghai Key Laboratory of Mechanics in Energy Engineering, Shanghai Institute of Applied Mathematics and Mechanics, School of Mechanics and Engineering Science, Shanghai University, Shanghai 200072, People's Republic of China}

\author{Run-Sen Zhang}
    \affiliation{Shanghai Key Laboratory of Mechanics in Energy Engineering, Shanghai Institute of Applied Mathematics and Mechanics, School of Mechanics and Engineering Science, Shanghai University, Shanghai 200072, People's Republic of China}

\author{Bing-Shen Wang}
    \affiliation{State Key Laboratory of Semiconductor Superlattice and Microstructure and Institute of Semiconductor, Chinese Academy of Sciences, Beijing 100083, China}

\date{\today}
\begin{abstract}

Zirconia is well-known for plenty of important morphologys with Zr coordination varying from sixfold in the octagonal phase to eightfold in the cubic or tetragonal phase. The development of empirical potentials to describe these zirconia morphologys is an important issue but a long-standing challenge, which becomes a bottleneck for the theoretical investigation of large zirconia structures. In contrast to the standard core-shell model, we develop a new potential for zirconia through the combination of long-range Coulomb interaction and bond order Tersoff model. The bond order characteristic of the Tersoff potential enables it to be well suited for the description of these zirconia morphologys with different coordination numbers. In particular, the complex monoclinic phase with two inequivalent oxygens, that is difficult to be described by most existing empirical potentials, can be well captured by this newly developed potential. It is shown that this potential can provide reasonable predictions for most static and dynamic properties of various zirconia morphologys. Besides its clear physical essence, this potential is at least one order faster than core-shell based potentials in the molecule dynamics simulation, as it discards the concept of the ultralight shell that demands for an extremely small time step. We also provide potential scripts for the widely used packages GULP and LAMMPS.

\end{abstract}
\maketitle
\pagebreak

\section{Introduction}
Zirconia-based ceramics are important industrial materials of high-temperature stability and high strength.\cite{KisiEH1998kem} The yttria-stabilized zirconia serves as the thermal barrier coating material to protect substrate components from hot gases in turbines and engines.\cite{PadtureNP2002sci,WuC2010scts,LiS2018scts,LiS2019scts} Zirconia can be used as an oxygen sensor or high quality oxygen ion channels, while artificial diamonds can be produced based on the zirconia.\cite{KisiEH1998}

Due to its industrial and military importance, zirconia and zirconia-based materials have attracted intense global research interest. Lots of research findings for the zirconia-based materials are first discovered by experiments rather than the theory. For instance, the successful application of yttria-stabilize zirconia as thermal barrier coating material was achieved in the experiment,\cite{PadtureNP2002sci} while theoretical studies fall far behind the experimental achievements. With the development of the computer speed, more and more {\it ab initio} calculations have been performed to study various properties for the zirconia.\cite{JansenHJF1988pb,ParlinskiK1997prl,JomardG1999prb,KuwabaraA2005prb,SouvatzisP2008prb,WuH2015jac,LiCW2015prb} {\it Ab initio} calculations are of high accuracy, but they are also computationally expensive. Empirical potentials are desiable for studying systems of hugh number of degrees of freedom.

As long as reliable empirical potentials are the foundation for theoretical researches, significant efforts have been devoted to developing empirical potentials for zirconia. Zirconia is an ionic oxide, so its interction is dominated by the Coulomb interaction. As a classic treatment, the long-range attractive Coulomb interaction is ballanced by the short-range Born-Mayer repulsive interaction,\cite{LewisGV1985jpcssp} which origins in the Pauli repulsion from the overlap of electron density.\cite{BornM} Several parameter sets for the Born-Mayer model are available for zirconia in the existing literature.\cite{SchellingPK2001jacs,KiloM2003pccp,YangC2018prb}

The Coulomb and Born-Mayer interactions together can provide a basic description for some properties of cubic zirconia, but they can not describe the important tetragonal and monoclinic phases. The stability of the tetragonal phase is closely related to the instantaneously polarization of ions. A standard approach to describe the polarizable ion is to divide the ion into a pair of core and shell.\cite{DickBG1958pb} The core-shell model has been parameterized for zirconia in several works.\cite{DwivediA1990pma,WilsonM1996prb,LauKC2011jpcm} The effect of polariable charges can also be considered by introducing a phenomenological charge-dipole term.\cite{ShimojoF1992jpsj} Besides, the instability of the cubic phase and the resultant c-t transition can also be predicted by introducing additional Born-Mayer interactions among oxygen ions.\cite{SchellingPK2001jacs}

The monoclinic phase is even more complicated with two inequivalent oxygen ions.\cite{KisiEH1998kem} It was shown that the existence of two inequivalent oxygen ions originates in the charge redistribution between oxygen ions in the zirconia.\cite{SmirnovM2003prb} To correctly describe all zirconia morphologys simultaneously, one needs to combine the Coulomb interaction, the Born-Mayer potential, the core-shell model, and the charge redistribution effect.\cite{SmirnovM2003prb}

Some other empirical potentials are also available to describe the atomic interaction within zirconia, including the tight-binding model,\cite{FabrisS2000prb} the reactive force field model,\cite{DuinACT2008jpca} and the neural network model.\cite{WangC2018ms}

In contrast to the core-shell model, we are going to analyze the interaction by examing the bond order properties of the zirconia. As mentioned by Smirnov {\it et al},\cite{SmirnovM2003prb} dioxides with larger cations (Th, Ce, U) are stable in the fluoritelike lattices with eightfold cation coordination, while dioxides with smaller cations (Pb, Sn, Ti, W) are stable in the structures with sixfold cation coordination. Different from these dioxides, the coordination varies among different zirconia morphologys. The lowest-energy state is the monoclinic phase with sevenfold cation coordination. The cubic/tetragonal phase has eightfold cation coordination, while the octagonal phase has sixfold cation coordination. The energy order for these zirconia phases is monoclinic $<$ tetragonal $<$ cubic $<$ octagonal, which explicitly shows a strong correlation between energy and the coordination.\cite{KisiEH1998kem} In other words, the configuration of zirconia depends on the bond order of the Zr atom. Inspired by this bond order dependence, we believe that the atomic interaction within zirconia may be described by bond order empirical potentials. The Tersoff model is a widely used empirical potential that possesses an explicit bond order dependence in its functional form.\cite{TersoffJ1} We thus propose to combine the Coulomb interaction and the Tersoff potential to describe the atomic interaction for zirconia.

In this paper, we suggest to describe the atomic interaction within zirconia by the Coulomb and Tersoff (CT) potential. The Tersoff potential captures different zirconia morphologys in terms of the bond order of Zr atoms, and the traditional core shell conception is avoided. As a consequence, the CT potential is at least one order faster than the standard core-shell based empirical potentials, while a clear physical essence is maintained. By taking advantage of the bond order property in the Tersoff potential, the CT potential is able to predict correct energy order for various zirconia morphologys, and in particular with the monoclinic phase as the lowest-energy morphology. The instability of the cubic zirconia is also predicted by the CT potential. We apply the CT potential to systematically investigate a series of static and dynamic properties for different zirconia morphologys.

\section{Structure}

\begin{figure*}[tb]
  \begin{center}
    \scalebox{0.8}[0.8]{\includegraphics[width=\textwidth]{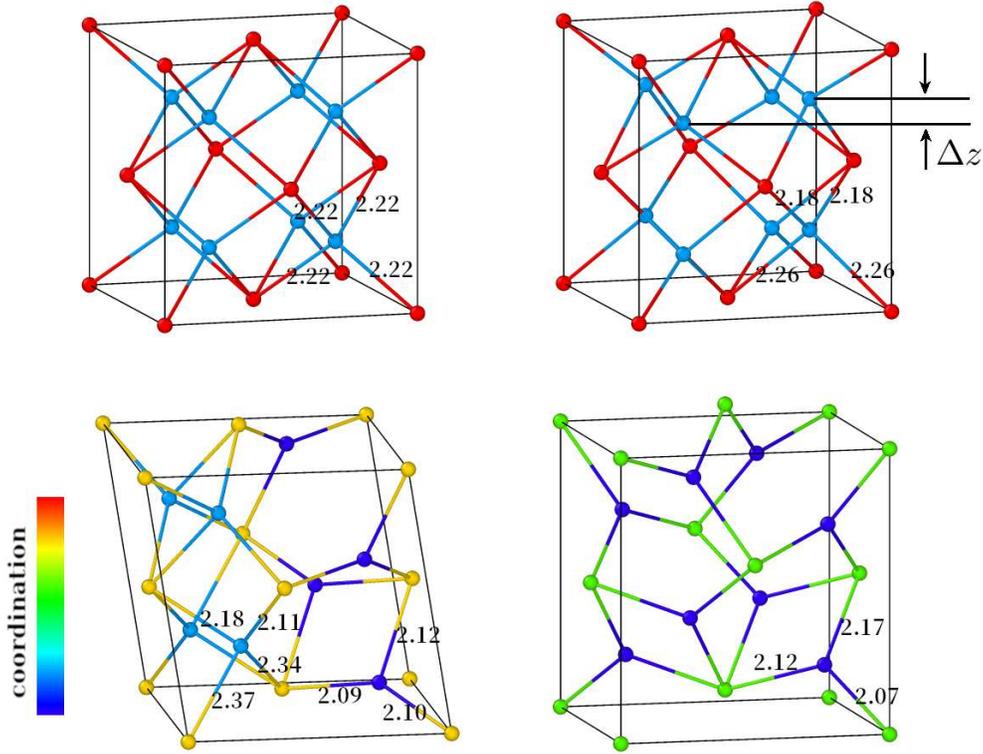}}
  \end{center}
  \caption{(Color online) Symmetric unit cells illustrate the atomic configuration for the (a) cubic, (b) tetragonal, (c) monoclinic, and (d) octagonal ZrO$_2$. The Zr-O bond length is also displayed. The colorbar is with respective to the coordination of each atom. Two inequivalent oxygen atoms, O$^{I}$ with 3 bonds and O$^{II}$ with 4 bonds, are clearly demonstrated for the monoclinic ZrO$_2$ in (c).}
  \label{fig_cfg}
\end{figure*}

Lots of zirconia morphologys have been observed in the experiment or discussed theoretically.\cite{KisiEH1998kem} The present work focuses on these four most studied zirconia morphologys shown in Fig.~\ref{fig_cfg}, including the cubic, tetrogonal, monoclinic, and octagonal phases. In the cubic zirconia, Zr atoms take the FCC lattice sites while oxygen atoms are in the tetrahedral position. Each Zr atom is coordinated by eight oxygen atoms in a symmetric manner. In the tetragonal zirconia, these eight oxygen atoms around the Zr atom are divided into two groups and relatively shifted for $\Delta z$ along one principal axis, while the symmetric unit cell is elongated along this principal axis. The monoclinic phase in Fig.~\ref{fig_cfg}~(c) has a more complex configuration with a monoclinic lattice. The Zr atom has a sevenfold coordination. There are two inequivalent oxygen atoms, one with threefold coordination while the other with fourfold coordination. Fig.~\ref{fig_cfg}~(d) shows an octagonal phase, in which Zr atoms have six coordination while oxygen atoms have threefold coordination. The octagonal phase is not observed in the experiment, so this phase shall have higher energy than other zirconia morphologys.

\section{Potential model}
\subsection{Born-Mayer potential}

\begin{figure}[tb]
  \begin{center}
    \scalebox{1}[1]{\includegraphics[width=8cm]{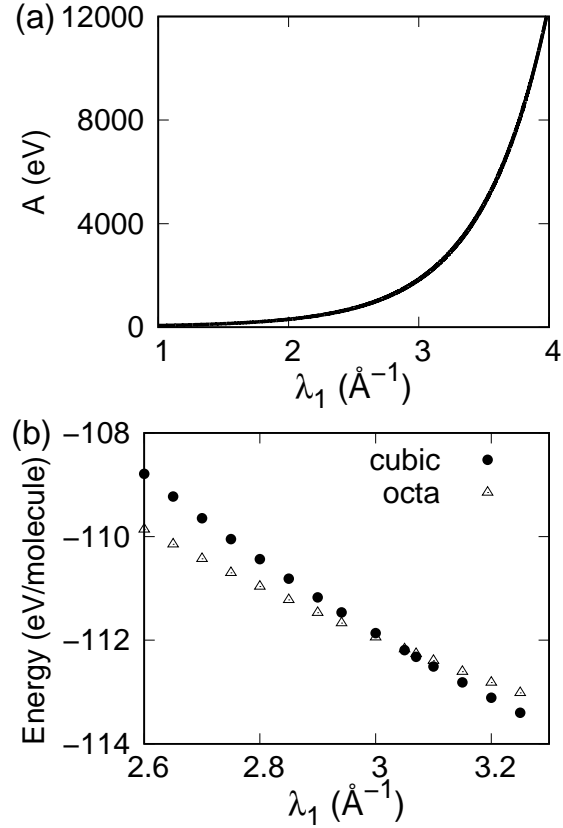}}
  \end{center}
  \caption{The constraint between parameters of the repulsive Born-Mayer interaction. (a) The constrain relationship between $A$ and $\lambda_1$. (b) The energy for the cubic and octagonal ZrO$_2$ for different sets of parameters ($A$, $\lambda_1$) obeying the constraint.}
  \label{fig_constraint}
\end{figure}

The interaction within ZrO$_{2}$ is usually described by the long-range attractive Coulomb interaction and the short-range repulsive Born-Mayer interaction\cite{BornM}
\begin{eqnarray}
V_{ij} & = & \frac{Q_{i}Q_{j}}{r_{ij}}+Ae^{-\lambda_{1}r_{ij}},
\label{eq_bm}
\end{eqnarray}
where $r_{ij}$ is the distance between atoms i and j. The first term is the Coulomb interaction between charges $Q_{i}$ and $Q_{j}$. The second term is the Born-Mayer interaction
with two parameters $A$ and $\lambda_{1}$. It has been shown that there is a constraint relation between these two parameters,\cite{SmirnovM2003prb} which can be obtained as follows. The force of the repulsive term is
\begin{eqnarray}
F & = & A\lambda_{1}e^{-\lambda_{1}r}.
\label{eq_fbm}
\end{eqnarray}
Request that the Zr-O bond length of $r_{0}=2.2$~{\AA} to be the same for cubic ZrO$_{2}$ with different sets of parameters $A$ and $\lambda_{1}$, so the force from the Coulomb interaction keeps unchanged. As a result, the same force ($F_{0}$) is obtained from the repulsive interaction with different sets of parameters $A$ and $\lambda_{1}$. The following constraint relationship between $A$ and $\lambda_{1}$ can thus be obtained
\begin{eqnarray}
F_{0} & = & A\lambda_{1}e^{-\lambda_{1}r_{0}}.
\label{eq_constraint}
\end{eqnarray}
This constraint relation is plotted in Fig.~\ref{fig_constraint}~(a).

For each set of parameters ($A$, $\lambda_{1}$), the energy for the cubic and octagonal ZrO$_{2}$ are compared in Fig.~\ref{fig_constraint}~(b). It is reasonable to guarantee that the octagonal ZrO$_{2}$ has higher energy than the cubic ZrO$_{2}$, as the octagonal ZrO$_{2}$ is not observed in experimental samples. A reasonable parameter set is $\lambda_{1}=3.05$~{\AA}, and $A=2023.6003$~{eV} according to the constraint relationship.

The Coulomb attractive interaction plus the Born-Mayer repulsive interaction captures the fundamental ionic characteristic features of the ZrO$_{2}$. As a consequence, several phases for the ZrO$_{2}$ can be obtained by structure relaxation with this potential, including the cubic, octagonal, and monoclinic phases. The tetragonal phase can also be obtained by introducing additional Born-Mayer repulsive interaction among oxygen ions.\cite{SchellingPK2001jacs} However, with the Coulomb and Born-Mayer interactions, it is impossible to obtain a correct order for the energy of different phases, as this potential does not contain the underlying mechanism for the stability of the tetragonal and monoclinic phases. Many works reveal that the polarization of the oxygen ion is the physical mechanism driving the c-t phase transition;\cite{DwivediA1990pma,WilsonM1996prb,LauKC2011jpcm} i.e., the polarization of the oxygen ion stabilizes the tetragonal phase of ZrO$_{2}$. The core-shell model is usually adopted to describe polarizable ions. For the monoclinic phase, some works proposed that the charge redistribution over these two inequivalent oxygen ions is the key mechanism to stabilize the monoclinic ZrO$_{2}$, and a variable charge model is developed to describe the stability of the monoclinic ZrO$_{2}$.\cite{SmirnovM2003prb}

\subsection{Tersoff potential}
In contrast to the core-shell model and the variable charge model, we suggest to describe zirconia morphologys by the CT potential that combines the Coulomb interaction and the Tersoff potential. The most significant characteristic of the Tersoff potential is its bond order property, i.e., the strength of each bond depends on its chemical environment. In particular, the bond strength depends explicitly on the coordination number of these two atoms forming this bond. As a result, the energy of each atom is dependent on its coordination. Recalling that Zr atoms have varying coordination number in different zirconia morphologys, the Tersoff potential is rather suitable in describing the interaction within zirconia.

\begin{table}
\caption{Parameters for the Coulomb interaction in zirconia.}
\label{tab_coulomb}
\begin{tabular}{@{\extracolsep{\fill}}|c|c|c|c|}
\hline 
$Q_{\rm Zr}$ & $Q_{\rm O}$ & $\alpha$ ({\AA}$^{-1}$) & cut off ({\AA})\tabularnewline
\hline 
\hline 
3.8 & -1.9 & 0.3 & 10.0\tabularnewline
\hline 
\end{tabular}
\end{table}

The CT potential takes the following form,
\begin{eqnarray}
V_{ij} & = & \frac{Q_{i}Q_{j}}{r_{ij}}+V_{ij}^{t},
\label{eq_ct}
\end{eqnarray}
where the first term is the standard Coulomb interaction. Parameters related to the Coulomb interaction are listed in Tab.~\ref{tab_coulomb}. The summation of the long-range electrostatic interaction is done by the truncation-based summation approach initially proposed by Wolf et al. in 1999\cite{WolfD} and further developed by Fennell and Gezelter in 2006.\cite{FennellCJ} We have chosen the damping parameter $\alpha=0.3$~{\AA$^{-1}$} and the cut-off $r_{c}=10.0$~{\AA}, which have been used in several previous works.\cite{DaiS2011jap,AgrawalR}

The second term in Eq.~(\ref{eq_ct}) is the Tersoff potential. This potential was first proposed by Tersoff in 1986,\cite{TersoffJ1} modified in 1988,\cite{TersoffJ3} and then generalized to multi-component system in 1989.\cite{TersoffJ5} There are some minor differences in notations of different versions. The present work uses the following functional form for the Tersoff potential,
\begin{eqnarray}
V_{ij}^{t} & = & f_{C}\left(r_{ij}\right)\left[f_{R}\left(r_{ij}\right)+b_{ij}f_{A}\left(r_{ij}\right)\right].
\label{eq_tersoff}
\end{eqnarray}
The cut-off function is
\begin{eqnarray}
f_{C}\left(r\right) & = & \begin{cases}
1, & r<R\\
\frac{1}{2}+\frac{1}{2}\cos\left(\pi\frac{r-R}{S-R}\right), & R<r<S\\
0, & r>S.
\end{cases}
\label{eq_fc}
\end{eqnarray}
The repulsive and attractive terms are
\begin{eqnarray}
f_{R}\left(r_{ij}\right) & = & Ae^{-\lambda_{1}r_{ij}};\\
f_{A}\left(r_{ij}\right) & = & -Be^{-\lambda_{2}r_{ij}}.
\end{eqnarray}
It shoud be noted that the repulsive term in the Tersoff potential is exactly the same as the Born-Mayer potential. Hence, the values for parameters $A$ and $\lambda_{1}$ in the Tersoff potential are set to be the same as that for the Born-Mayer potential, i.e., $\lambda_{1}=3.05$~{\AA}, and $A=2023.6003$eV.

Following the Morse potential,\cite{MorsePM1929pb} parameter $\lambda_{2}$ in the attractive term can be set by $\lambda_{2}=\lambda_{1}/2$, i.e., $\lambda_{2}=1.525$~{\AA}. To determine the energy parameter $B$, we point out the fact that both Coulomb interaction and the $f_{A}$ term in the Tersoff potential are attractive interactions. The ionic bond model is most suitable for ionic crystals consisted by the atoms from columns I and VII in the periodic table. It is because the energy of valence electron in the metallic atom is much higher than that of the chlorine-like atom, so the valence electron transfer is nearly complete while the electron coupling is only a small perturbation. Zirconia is consisted of the transition metal from column II and the oxygen from column VI, both of which move to the center of the periodic table. The energy difference between the valence electron of Zr and O atoms is smaller, so the valence electron transfer is slightly weakened. As a result, the covalent component will increase while the ionicity still dominates, which will be reflected by the reduction of the ionic charges to the effective charges.\cite{HarrisonWA2004} Indeed, we find that the charges of the Zr cations and the O ions are slightly reduced from their normal values to +3.8 and -1.9, which indicates that a small portion of the ionic interaction is replaced by the covalent interaction. The fitted parameter, $B=17.3376$~{eV}, is about two orders smaller than the parameter A, which further confirms that the interaction is mostly ionic.

The characteristic feature of the Tersoff potential is the bond order term
\begin{eqnarray}
b_{ij} & = & \left(1+\beta^{n}\zeta_{ij}^{n}\right)^{-\frac{1}{2n}}.
\label{eq_bij}
\end{eqnarray}
The effective coordination $\zeta_{ij}$ includes the local environment effect through the following expression
\begin{eqnarray}
\zeta_{ij} & = & \sum_{k\not=i,j}f_{c}\left(r_{ik}\right)g\left(\theta_{ijk}\right)e^{\lambda_{3}^{m}\left(r_{ij}-r_{ik}\right)^{m}},
\label{eq_zeta}
\end{eqnarray}
where the summation $\sum_{k}$ is over other bonds i-k around atom i. The coordination for atom i is in close relation to the quantity $\zeta_{ij}$, which is regarded as the effective coordination for atom i. The three-body term is
\begin{eqnarray}
g\left(\theta_{ijk}\right) & = & 1+\frac{c^{2}}{d^{2}}-\frac{c^{2}}{d^{2}+\left(h-\cos\theta_{ijk}\right)^{2}}.
\label{eq_thetaijk}
\end{eqnarray}

\begin{figure}[tb]
  \begin{center}
    \scalebox{1}[1]{\includegraphics[width=8cm]{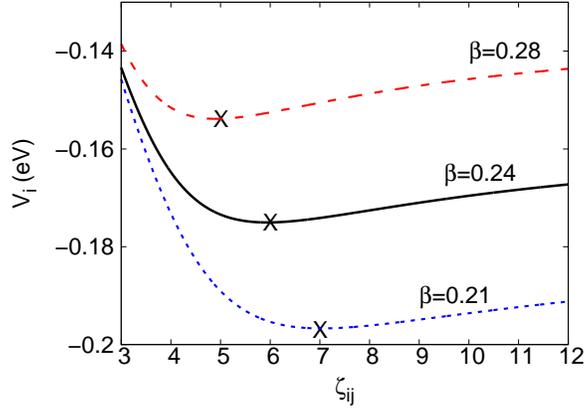}}
  \end{center}
  \caption{The dependence of the atomic energy (V$_i$) on the effective coordination ($\zeta_{ij}$) from the Tersoff potential with $n=5.0$ and different $\beta$. Note that the energy is minimized around $\zeta_{ij}=6$ (the actual coordination is 7) for $\beta=0.24$.}
  \label{fig_tersoff-vi}
\end{figure}

Before presenting parameters for the Tersoff potential, we discuss the suitability of the Tersoff potential in describing the monoclinic phase by taking advantage of its bond order characteristic. Let's consider a simple situation with $g(\cos\theta_{ijk})=1$ and $\lambda_{3}=0.0$ in Eq.~(\ref{eq_zeta}), so the actual coordination for atom i is $\zeta_{ij}+1$. Due to this close relationship between the quantity $\zeta_{ij}$ and the coordination, the quantity $\zeta_{ij}$ is usually called the effective coordination. The energy minimum position $r_{m}$ is determined by
\begin{eqnarray}
\frac{\partial V_{ij}^{t}}{\partial r}|_{r=r_{m}} & = & 0,
\end{eqnarray}
which gives
\begin{eqnarray}
r_{m} & = & \frac{1}{\lambda_{1}-\lambda_{2}}\ln\left(\frac{A\lambda_{1}}{bB\lambda_{2}}\right).
\end{eqnarray}
The corresponding energy minimum is
\begin{eqnarray}
V_{ij}^{m} & = & Ae^{-\lambda_{1}r_{m}}-bBe^{-\lambda_{2}r_{m}}.
\end{eqnarray}
Assuming all bonds around atom i have the same energy, the total energy for atom i can be obtained
\begin{eqnarray}
V_{i}(\zeta) & = & V_{m}\times\left(\zeta+1\right)\\
 & = & \left(Ae^{-\lambda_{1}r_{m}}-bBe^{-\lambda_{2}r_{m}}\right)\times\left(\zeta+1\right).
\label{eq_vi}
\end{eqnarray}
The subscripts i and j have been omitted here. Note that the quantity $b$ is also a function of $\zeta$ as can be seen in Eq.~(\ref{eq_bij}). For a given value of ($n$, $\beta$), the energy $V_{i}(\zeta)$ is an explicit function of the effective coordination $\zeta$. Fig.~\ref{fig_tersoff-vi} displays the function $V_{i}(\zeta)$, which shows that the minimum of $V_{i}(\zeta)$ with respective to $\zeta$ can be well manipulated by tuning parameters ($n$, $\beta$). Particularly, the energy $V_{i}(\zeta)$ has its minimum value at $\zeta=6$ (i.e., actual coordination of 7) for n=5 and $\beta=0.24$. We thus demonstrate that the bond order characteristic of the Tersoff potential is able to ensure the monoclinic phase (with sevenfold coordination for Zr) to be the lowest-energy configuration among all zirconia morphologys. We have thus shown that the stability of the monoclinic zirconia among different morphologys (with different Zr coordination) can be well described by the Tersoff potential using its bond order property. The parameters ($n$, $\beta$) for the Zr atom are primarily determined according to Fig.~\ref{fig_tersoff-vi}. It should be noted that in zirconia morphologys with inequivalent bond lengths, these bonds around an atom will have different energy, so the assumption in Eq.~(\ref{eq_vi}) does not hold in this situation. Consequently, the final parameters ($n$, $\beta$) are slightly deviated from the ideal values in Fig.~\ref{fig_tersoff-vi}.

We emphasize that there are both repulsive and attractive terms in the Tersoff potential. The repulsive term in the Tersoff potential is exactly the same as the usual Born-Mayer potential, both of which describe the Pauli repulsion due to the overlap of electron density. The novelty of the CT potential lies in the bond order dependent attractive term in the Tersoff potential. The bond order attractive term is of small fraction in the whole CT potential, but this is the kernel ingredient that provides good descriptions for both tetragonal and monoclinic morphologys.

\begin{table}
\caption{Parameters for the Tersoff potential of zirconia.}
\label{tab_tersoff}
\begin{tabular}{@{\extracolsep{\fill}}|c|c|c|c|c|c|c|}
\multicolumn{7}{c}{two-body}\tabularnewline
\hline 
A (eV) & B (eV) & $\lambda_{1}$ ({\AA}$^{-1}$) & $\lambda_{2}$ ({\AA}$^{-1}$) & R ({\AA}) & \multicolumn{2}{c|}{S ({\AA})}\tabularnewline
\hline 
2023.6003 & 17.3376 & 3.0500 & 1.5250 & 2.85 & \multicolumn{2}{c|}{3.15}\tabularnewline
\hline 
\multicolumn{7}{c}{three-body for Zr}\tabularnewline
\hline 
m & $\lambda_{3}$ ({\AA}$^{-1}$) & $\beta$ & n & c & d & h\tabularnewline
\hline 
3 & 0 & 0.2403 & 5.0062 & 0.0 & 0.0 & 0.0\tabularnewline
\hline 
\multicolumn{7}{c}{three-body for O}\tabularnewline
\hline 
m & $\lambda_{3}$ ({\AA}$^{-1}$) & $\beta$ & n & c & d & h\tabularnewline
\hline 
3 & 0 & 0.0601 & 2.2611 & 2.0204 & 0.1093 & -0.4112\tabularnewline
\hline 
\end{tabular}
\end{table}

Parameters for the Tersoff potential of ZrO$_{2}$ are listed in Tab.~\ref{tab_tersoff}. Parameters in the three-body term $g(\cos\theta)$ are fitted to the correct energy order of different zirconia morphologys. The Tersoff potential files for GULP\cite{gulp} and LAMMPS\cite{PlimptonSJ} are available from the personal website of the corresponding author (jiangjinwu.org).

\section{Results and discussions}

We have developed the CT potential for the zircornia in the above. The rest of this paper is devoted to applying this CT potential to study some typical properties for different zirconia morphologys, and compare these results with available experiments or {\it ab initio} calculations.

\subsection{Monoclinic phase}

\begin{table}
\caption{Structural properties for the monoclinic ZrO$_{2}$. The second line lists the lattice constants. $\beta$ is the tilting angle. Length is in the unit of {\AA}. Angle is in the unit of degree.}
\label{tab_mzro2}
\begin{tabular}{@{\extracolsep{\fill}}|c|c|c|c|c|}
\hline 
 & exp.\cite{KisiEH1998} & {\it ab initio}\cite{JomardG1999prb} & VCM\cite{SmirnovM2003prb} & CT, this work\tabularnewline
\hline 
\hline 
a,b,c & 5.145, 5.210, 5.312 & 5.242, 5.305, 5.410 & 5.167, 5.157, 5.323 & 5.4238, 4.9774, 5.3329\tabularnewline
\hline 
$\beta$ & 99.2 & 99.23 & 97.2 & 95.2\tabularnewline
\hline 
Zr  & 0.2751, 0.0404, 0.2081 & 0.2765, 0.0421, 0.2090 & 0.2806, 0.0176, 0.2194 & 0.2777, 0.0414, 0.2102\tabularnewline
\hline 
O$^{I}$  & 0.0770, 0.3351, 0.3437 & 0.071, 0.337, 0.342 & 0.0468, 0.2973, 0.3812 & 0.0755, 0.3275, 0.3960\tabularnewline
\hline 
O$^{II}$  & 0.5480, 0.2425, 0.5250 & 0.550, 0.242, 0.521 & 0.5303, 0.2470, 0.5163 & 0.5391, 0.2686, 0.5205\tabularnewline
\hline 
Zr-O$^{I}$ & 2.0371, 2.0838, 2.1391 & 2.0915, 2.1017, 2.1972 & 2.0533, 2.0974, 2.2620 & 2.0870, 2.0995, 2.1238\tabularnewline
\hline 
\multirow{2}{*}{Zr-O$^{II}$} & 2.1446, 2.1548, & 2.1923, 2.2067, & 2.0969, 2.1412, & 2.1104, 2.1795,\tabularnewline
 & 2.2548,2.2782 & 2.2919, 2.2963 & 2.2117, 2.2828 & 2.3388, 2.3669\tabularnewline
\hline 
\end{tabular}
\end{table}

\begin{table}
\caption{Static properties for zirconia of different phases.}
\label{tab_static}
\begin{tabular}{@{\extracolsep{\fill}}|c|c|c|c|c|c|c|c|}
\hline 
\multicolumn{3}{|c|}{Volume} & c-t & Imaginary & \multicolumn{2}{c|}{Energy barrier} & Method and\tabularnewline
\multicolumn{3}{|c|}{(\AA$^{3}$/molecule)} & distortion & mode (cm$^{-1}$) & \multicolumn{2}{c|}{(meV/molecule)} & reference\tabularnewline
\hline 
\hline 
$V_{c}$ & $V_{t}$ & $V_{m}$ & $\Delta z$ & $\omega_{X_2^-}$ & $\Delta E_{ct}$ & $\Delta E_{tm}$ & \tabularnewline
\hline 
33.1 & 33.5 & 35.0 & 0.033 &  & 31.2 & 74.9 & {\it ab initio}\cite{ChristensenA1998prb}\tabularnewline
\hline 
32.9 & 33.7 & 35.1 & 0.06 &  & 56.2 & 62.4 & exp.\cite{KisiEH1998}\tabularnewline
\hline 
34.3 & 35.9 & 37.1 & 0.050 &  & 81.1 & 99.9 & {\it ab initio}\cite{JomardG1999prb}\tabularnewline
\hline 
32.7 & 33.1 & 35.2 & 0.039 & i120 & 16.2 & 15.6 & VCM\cite{SmirnovM2003prb}\tabularnewline
\hline 
33.66 & 33.70 & 35.84 & 0.013 & i155.8 & 3.05 & 80.27 & CT, this work\tabularnewline
\hline 
\end{tabular}
\end{table}

The CT potential includes both the long-range Coulomb interaction and the bond order Tersoff potential. The Tersoff potential has been implemented in most lattice dynamics or molecule dynamics (MD) simulation packages, like GULP\cite{gulp} and LAMMPS\cite{PlimptonSJ}. In the present work, the GULP\cite{gulp} package is used to calculate static properties for different zirconia morphologys, including structural properties, the energy barrier, the phonon dispersion, the Young's modulus, and the Poisson's ratio.

Table~\ref{tab_mzro2} shows that structural properties for the monoclinic zirconia calculated by the present CT potential are in good agreement with experiments or {\it ab initio} results, which implies that the CT potential can successfully describe the monoclinic phase. Furthermore, Tab.~\ref{tab_static} shows that the monoclinic phase has the lowest energy among all zirconia morphologys. The value of the energy barrier $\Delta E_{tm}=80.27$~{meV/molecule} between the tetragonal phase and the monoclinic phase is in good agreement with the experiments\cite{KisiEH1998} or the {\it ab initio} results.\cite{ChristensenA1998prb,JomardG1999prb} The volume of the monoclinic zirconia is obviously larger than the cubic and tetragonal phases, which agrees quite well with previous works. These promising results confirm that the present CT potential indeed predicts the monoclinic zirconia to be the most stable phase among different zirconia morphologys.

\subsection{Tetrogonal phase}

\begin{figure}[tb]
  \begin{center}
    \scalebox{1}[1]{\includegraphics[width=8cm]{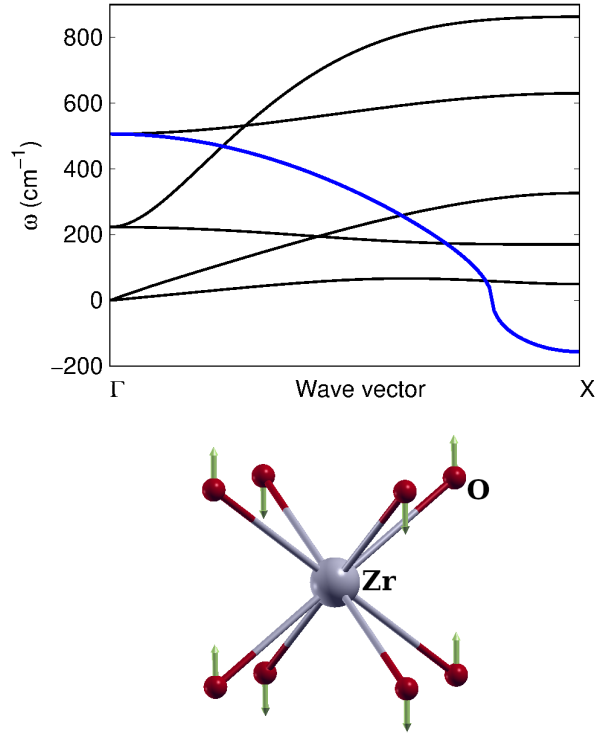}}
  \end{center}
  \caption{Phonon dispersion for cubic ZrO$_2$, calculated with the primitive unit cell of one Zr atom and two O atoms. An imaginary branch (blue online) exists around the boundary X of the Brillouin zone. The bottom inset displays the vibrational morphology of the phonon mode at the X point, i.e., the X$_2^{-}$ mode.}
  \label{fig_phonon}
\end{figure}

\begin{figure}[tb]
  \begin{center}
    \scalebox{1}[1]{\includegraphics[width=8cm]{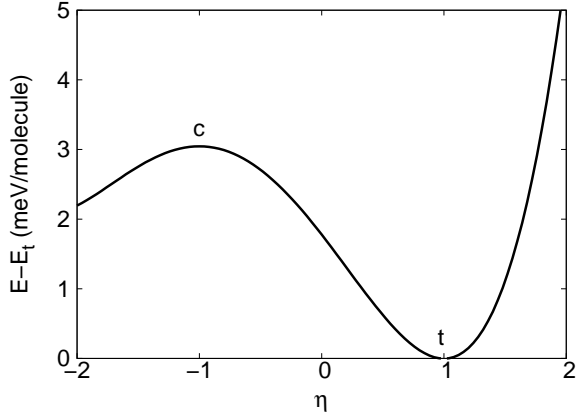}}
  \end{center}
  \caption{The energy for structures evolving from cubic to tetragonal ZrO$_2$. An arbitrary structure corresponding to $\eta$ is described by $\vec{R}_{\eta}=\frac{1-\eta}{2}\vec{R}_c+\frac{1+\eta}{2}\vec{R}_t$, where $\vec{R}_{c}$ and $\vec{R}_{t}$ are structures for the cubic and tetragonal phases, respectively.}
  \label{fig_evolve}
\end{figure}

While the success of the CT potential in describing the monoclinic phase is as expected considering the bond order feature of the Tersoff potential, we point out that the Tersoff potential does not have any specific features to guarantee the transition of the cubic phase into the more stable tetragonal phase. However, it is quite interesting that the present CT potential can also predict such transition. To illustrate this fact, we calculate the phonon dispersion for the cubic zirconia in Fig.~\ref{fig_phonon}. A primitive unit cell containing one Zr atom and two oxygen atoms is used for this calculation. The phonon dispersion is calculated with GULP.\cite{gulp} These phonon branches show similar behaviors as results from {\it ab initio} calculations.\cite{ParlinskiK1997prl,KuwabaraA2005prb,SouvatzisP2008prb,WuH2015jac} The most significant feature in the phonon dispersion is the softening of one optical branch, which eventually becomes imaginary for wave vectors around the X point in the Brillouin zone. Note a convention that an imaginary value will be shown as a negative value for the frequency in phonon dispersions like Fig.~\ref{fig_phonon}.

The vibratonal morphology corresponding to the imaginary mode at the X point, i.e., X$_2^-$ mode, is shown in the bottom inset of Fig.~\ref{fig_phonon}. This figure is plotted with the XCRYSDEN package.\cite{xcrysden} The Zr atom does not vibrate, while these eight surrounding oxygen atoms are divided into two groups. These two groups of oxygen atoms vibrate in an opposite direction. The cubic phase will be distorted into the tetragonal phase by deformation following the vibrational morphology of the X$_2^-$ mode. The X$_2^-$ mode is the origin for the transition of the cubic zirconia into the tetragonal zirconia.

The imaginary mode at the X point can be attributed to the instantaneous polarization of the Zr cation or the oxygen ions, which can be described by the core-shell model. Adding additional Born-Mayer interactions among oxygen ions can also provide a correct prediction for this imaginary mode. Here, we have provided a third solution for the X$_2^-$ imaginary mode, i.e., by the bond order Tersoff potential.

To further explore the relationship between the cubic and tetragonal phases, we examine the evolution of the structure from the cubic phase to the tetragonal phase. We introduce a parameter $\eta$ to evolve the structure according to the expression $\vec{R}_{\eta}=\frac{1-\eta}{2}\vec{R}_c+\frac{1+\eta}{2}\vec{R}_t$, where $\vec{R}_{c}$ and $\vec{R}_{t}$ represent the structure of cubic and tetragonal phases, respectively. The structure with $\eta=-1$ is the cubic phase, while the structure with $\eta=+1$ corresponds to the tetragonal phase. For an arbitrary $\eta$, the structure $\vec{R}_{\eta}$ is constructed based on the cubic and tetragonal configurations. Fig.~\ref{fig_evolve} shows the evolution of the energy for the structure by varying $\eta$. The cubic phase locates at a local maximum energy position, while the tetragonal phase is at a local minimum energy position. As a result, the cubic phase is unstable and will transform into the tetragonal phase.

\subsection{Static and mechanical properties}

\begin{table}
\caption{The Young's modulus (Y, in the unit of GPa) and the Poisson's ratio ($\nu$) for ZrO$_2$ of different phases. The Young's modulus is anisotropic with three different values along the x, y, and z directions.}
\label{tab_young}
\begin{tabular}{@{\extracolsep{\fill}}|c|c|c|c|c|c|}
\hline 
\multicolumn{2}{|c|}{c} & \multicolumn{2}{c|}{t} & \multicolumn{2}{c|}{m}\tabularnewline
\hline 
Y & $\nu$ & Y & $\nu$ & Y & $\nu$\tabularnewline
\hline 
\hline 
479.7 & 0.243 & $\begin{array}{c}
464.5\\
464.5\\
382.7
\end{array}$ & $\begin{array}{ccc}
/ & 0.234 & 0.251\\
0.234 & / & 0.251\\
0.305 & 0.305 & /
\end{array}$ & $\begin{array}{c}
633.3\\
392.5\\
430.1
\end{array}$ & $\begin{array}{ccc}
/ & 0.247 & 0.119\\
0.399 & / & 0.280\\
0.175 & 0.256 & /
\end{array}$\tabularnewline
\hline 
\end{tabular}
\end{table}

\begin{table}
\caption{Properties for the octagonal ZrO2 predicted by the CT potential in this work.}
\label{tab_ozro2}
\begin{tabular}{@{\extracolsep{\fill}}|c|c|c|c|c|c|}
\hline 
a, b, c & Zr-O bond & Volume & Energy barrier $E_{O}-E_{M}$ & Young's modulus & Poisson's\tabularnewline
(\AA) & (\AA) & (\AA/molecule) & meV/molecule & (GPa) & ratio\tabularnewline
\hline 
\hline 
$\begin{array}{c}
5.339\\
5.010\\
5.584
\end{array}$ & 2.1225 & 37.3 & 680.54 & $\begin{array}{c}
306.9\\
209.9\\
364.0
\end{array}$ & $\begin{array}{ccc}
/ & 0.360 & 0.249\\
0.526 & / & 0.526\\
0.210 & 0.300 & /
\end{array}$\tabularnewline
\hline 
\end{tabular}
\end{table}

Structural and energy properties for zirconia morphologys are listed in Tab.~\ref{tab_mzro2}. The magnitude for the frequency of the X$_2^-$ mode is comparable with previous calculations, which again verifies the instability of the cubic phase. The quantity $\Delta z$ as displayed in Fig.~\ref{fig_cfg} quantifies the magnitude of the c-t distortion. The $\Delta z$ from the present work is smaller than previous works,\cite{ChristensenA1998prb,KisiEH1998,JomardG1999prb,SmirnovM2003prb} which indicates that the c-t distortion is underestimated by the present CT potential. Due to the same reason, the energy barrier $\Delta E_{ct}$ between the cubic and tetragonal phases is underestimated by the present CT potential. These results illustrate that the present CT potential can only provide a qualitative description for the distortion of the cubic phase into the tetragonal phase. The actual magnitude for the distortion is underestimated within the CT potential.

The Young's modulus and the Poisson's ratio for different zirconia morphologys are compared in Tab.~\ref{tab_young}. We apply the present CT potential to predict properties for the octagonal zirconia in Tab.~\ref{tab_ozro2}. The energy of the octagonal zirconia is much higher than other zirconia morphologys, so the octagonal phase is quite unstable. This is consistent with the fact that the octagonal is never observed in the experiment.\cite{KisiEH1998kem}

\subsection{Dynamic properties}

\begin{figure}[tb]
  \begin{center}
    \scalebox{1}[1]{\includegraphics[width=8cm]{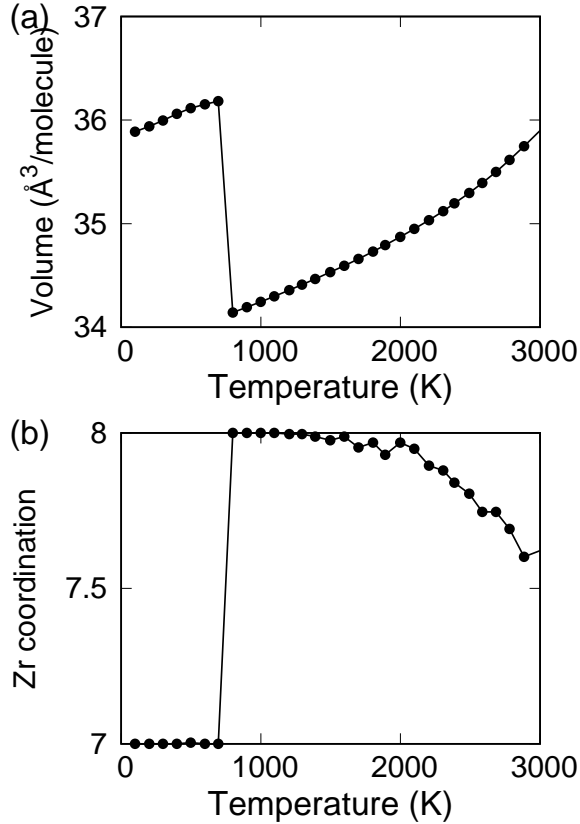}}
  \end{center}
  \caption{Phase transition for the monoclinic ZrO$_2$ by increasing temperature. (a) The temperature dependence for the volume of ZrO$_2$. (b) The temperature dependence for the Zr coordination. Note an obvious phase transition around 747~K.}
  \label{fig_melt}
\end{figure}

We now apply the CT potential for fMD simulations. MD simulations are performed using the publicly available simulation code LAMMPS.\cite{PlimptonSJ} The OVITO package is used for visualization of the MD snap shots.~\cite{ovito} The standard Newton equations of motion are integrated in time using the velocity Verlet algorithm with a time step of 1.0~{fs}. The structure has $2\times 2\times 2$ symmetric unit cells. A larger structure of $4\times 4\times 4$ symmetric unit cells has also been simulated and similar results are obtained. Periodic boundary conditions are apllied in all of the three directions in the present MD simulations.

\begin{figure*}[tb]
  \begin{center}
    \scalebox{1}[1]{\includegraphics[width=\textwidth]{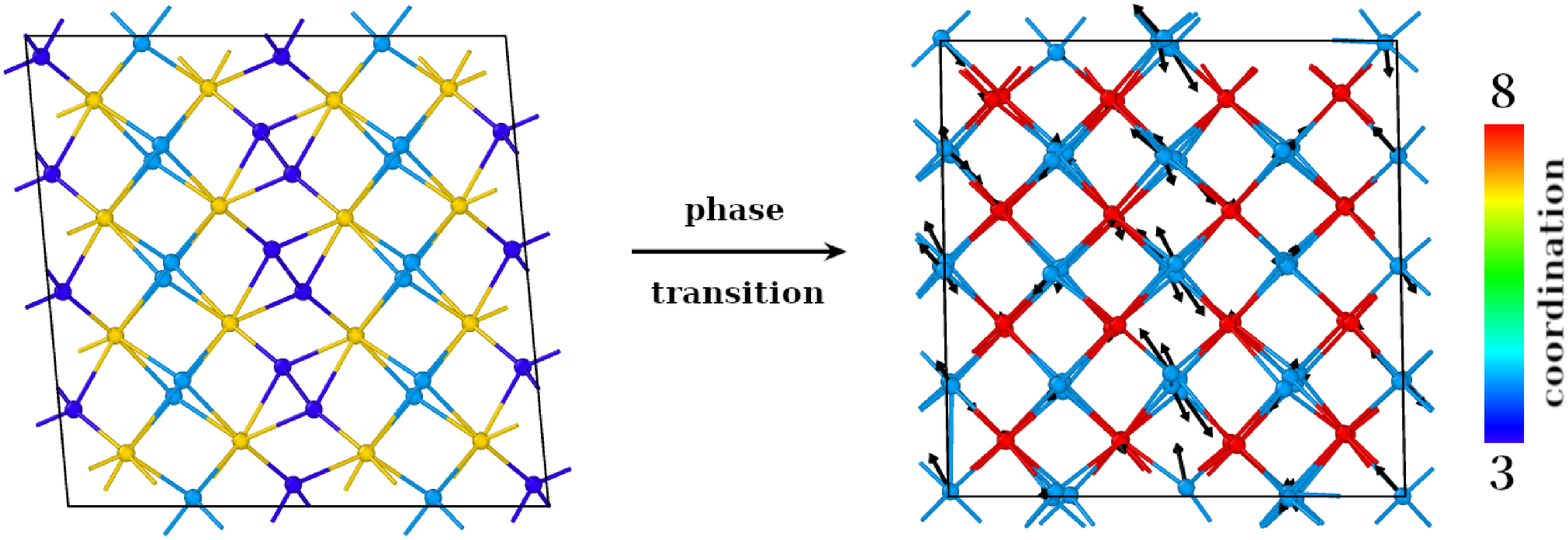}}
  \end{center}
  \caption{MD snapshots for the ZrO$_2$ just before and after the phase transition around 747~K. The arrow attached to each atom describes the atomic displacement induced by the phase transition. The colorbar is with respective to the coordination.}
  \label{fig_snapshot}
\end{figure*}
We simulate the heating process of the monoclinic zircornia. The evolution of the volume is plotted in Fig.~\ref{fig_melt}~(a) during this process. It shows a abrupt reduction in the volume around 747~K, which indicates possible m-t phase transition considering the smaller volume of the tetragonal phase. Fig.~\ref{fig_melt}~(b) shows that the Zr coordination increases from seven to eight after this phase transition, which further confirms the transition from the monoclinic phase (with sevenfold Zr coordination) into the tetragonal phase (with eightfold Zr coordination). Note that the decrease of the coordination in the high temperature range is due to the strong thermal vibration at high temperatures, where some neighboring oxygen atoms vibrate into a distance far from the Zr atom.

To explore more details of the structural transition, we show in Fig.~\ref{fig_snapshot} the atomic displacement caused by the phase transition. Note that the colorbar is with respective to the coordination number of each atom. It clearly displays the structure transition from the monoclinic phase into the tetragonal phase. The arrow on each atom represents its displacement induced by the phase transition. The oxygens O$^{I}$ with threefold coordination are seriously reconstructed by large displacements and are eventually deformed into fourfold coordination.

It should be noted that the critical temperature for the m-t transition predicted by the present CT potential is lower than the experimental value.\cite{KisiEH1998kem} Furthermore, there is no obvious t-c phase transition at higher temperatures, which shall occurs around 2377~K in the experiment.\cite{KisiEH1998kem} It is because the c-t distortion is underestimated by the present CT potential, so the c-t distortion can not take effect at the high temperature.

\section{Conclusion remarks}

Before conclusion, we address some general remarks on positive and negative properties of the present CT potential, to facilitate readers to decide whether the CT potential is suitable for their researches. The most significant feature of the CT potential is to use the bond order Tersoff potential to substitute the core-shell model. Both positive and negative properties of the CT potential are directly resultant from this substitution. We list these positive and negative properties as follows.

{\it Positive properties.} (1) The capability of the CT potential in describing the monoclinic phase is mainly owing to the bond order characteristic of the Tersoff potential, so the CT potential possesses a clear physical essence. (2) The CT potential predicts correct energy order for these four zircornia morphologys, with the monoclinic phase as the lowest-energy structure. (3) A clear m-t phase transition is observed in the MD simulation. (4) The CT potential is at least one order faster than the core-shell model in the MD simulation. It is because, to mimic an addibatic response of the shells to the cores, shells practically have very small mass and thus require ultra-small time step in MD simulations. (5) The Tersoff potential has been widely used in the computational community and has been implemented in most simulation packages, so the CT potential can be conveniently used.

{\it Negative properties.} (1) The magnitude of the c-t distortion is weaker than the experiment, and as a consequence the t-c phase transition is not observed in the MD simulation. (2) The polarization effect is treated effectively by the Tersoff potential without introducing shells, so the effect from the electric field on the polarization of the shells can not be simulated by the CT potential.

To summary, we have developed a potential by combining the Coulomb interaction and the Tersoff potential to describe the atomic interaction for zirconia. The bond order property of the Tersoff potential enables this potential to be very suitable in describing these well-known zirconia morphologys. More specifically, within this potential, the monoclinic zirconia is the most stable phase in the low temperature region. The cubic phase is not stable and will be spontaneously distorted into the tetragonal phase. The octagonal phase has the highest energy and shall not be observed in the experiment. These predictions agree quite well with the experiments or {\it ab initio} calculations. We also use this potential to predict various static or dynamic properties for the zirconia morphologys. The potential scripts for GULP and LAMMPS are available from the website of the corresponding author (jiangjinwu.org).

\textbf{Acknowledgment} The work is supported by the National Natural Science Foundation of China (NSFC) under Grant No. 11822206 and the Innovation Program of Shanghai Municipal Education Commission under Grant No. 2017-01-07-00-09-E00019.

%
\end{document}